# An Architecture for Modular Data Centers

James Hamilton
Microsoft Corporation
One Microsoft Way
Redmond, WA 98052
+1 (425)703-9972

JamesRH@microsoft.com

http://research.microsoft.com/~jamesrh/

## ABSTRACT
Large numbers of low-cost, low-reliability commodity components are rapidly replacing high-quality, mainframe-class systems in data centers. These commodity clusters are far less expensive than the systems they replace, but they can bring new administrative costs in addition to heat and power-density challenges. This proposal introduces a data center architecture based upon macro-modules of standard shipping containers that optimizes how server systems are acquired, administered, and later recycled.

## Categories and Subject Descriptors
K.6.4 [**Management of Computing and Information Systems**]: Systems Management – *data center design, architecture, high scale deployment, commodity server-side computing.*

## General Terms
Management, Economics, Experimentation.

## Keywords
Data centers, Google, liquid cooling, shipping container, Sun Microsystems, Rackable Systems, Microsoft, total cost of ownership.

## 1. INTRODUCTION
Internet-scale services built upon commodity computing clusters are more affordable than ever, and are becoming increasingly common. Drivers of this trend on the consumption side include the migration of on-premise applications to Software-as-a-Service (SaaS) providers, the increased use of commercial high performance computing, and the emergence of mega-scale consumer services.

Also accelerating this trend is the need for increased reliability and the absence of scheduled downtime when serving a world-wide audience. There is no "service window". Even highly reliable hardware fails to achieve five 9s, so redundant clusters must be used. Once the software has been written to run efficiently and mask failure over a redundant cluster, then much cheaper and less reliable hardware components can be used as the cluster building blocks without negatively impacting the reliability of the overall service.

Commodity systems substantially reduce the cost of server-side computing. However, they bring new design challenges, some technical and some not. The technical issues include power consumption, heat density limits, communications latencies, multi-thousand-system administration costs, and efficient failure management with large server counts. The natural tendency is towards building a small number of very large data centers and this is where some of the non-technical issues come to play. These include social and political constraints on data center location, in addition to taxation and economic factors. All diminish the appeal of the large, central data center model. Multiple smaller data centers, regionally located, could prove to be a competitive advantage.

To address these technical and non-technical challenges, we recommend a different granule of system purchase, deployment, and management. In what follows, we propose using a fully-populated shipping container as the data-center capitalization, management, and growth unit. We argue that this fundamental change in system packaging can drive order-of-magnitude cost reductions, and allow faster, more nimble deployments and upgrades.

## 2. COMMODITY DATA CENTER GROWTH DRIVERS
We're currently witnessing rapid expansion in the world-wide data center inventory and increasing system density in those centers. Three main factors are driving this growth: 1) the proliferation of high-scale Software as a Service (SaaS) providers, 2) the emerging importance of commercial high-performance computing and 3) rapid growth in the online consumer services sector fueled by the Internet and the success of the advertising-based revenue model.

## 2.1 Software as a Service

Developing a competitive services business is dependent upon the cost of providing that service. This is one factor that makes the SaaS model fundamentally different from packaged software. In the SaaS model, the cost of selling a unit of service is real and backed by physical assets; whereas, with packaged software, the marginal sales cost for a broadly distributed product are near zero. With SaaS, the customer is purchasing the aggregation of the software that implements the solution; the hardware that hosts the software; the data center that houses the hardware; the admin staff that takes care of the data centers and systems; and the customer support staff that takes care of the users. The SaaS model appeals to customers because the cost is a predictable operational expense that scales with business growth, rather than a fixed capital outlay made in advance. Just as very few companies currently process their own payroll, more and more are moving their internal systems to a SaaS model. IDC forecasts a continued 25% compound annual growth rate in SaaS [10].

## 2.2 Commercial High Performance Computing

Companies in most business areas are exploiting the rapidly falling price of computing and using information technology to better understand their customers; to understand and control their costs; and to add more value. Leaders in this area, such as Wal-Mart and Charles Schwab, have long depended upon software systems to reduce costs and improve services. And, as the price of server-side-computing continues to decline, these same technologies and techniques [2][15] are being exploited much more broadly.

## 2.3 Consumer Services

Google continues to be the showcase consumer of high-scale, commodity system resources, and remains a leader in mass deployment. In September of 2002, Google reported that it had 15,000 servers in six data centers [8]. In a 2005 paper, Stephen Arnold reported that Google had grown to over 100,000 systems in 15 data centers [1]. Recent estimates run in excess of 500,000 systems in over 30 data centers world-wide. At these scales, the infrastructure costs are enormous. Even small changes in acquisition, deployment and systems management costs will have a fundamental impact on the aggregated costs.

The cost of purchasing the individual systems that form the base building block of these clusters is fairly low at $1,000 to $3,000. The dominant cost is shipping, packaging, deploying, housing, powering, cooling, and replacing/repairing these systems. The architectural changes proposed in what follows are aimed at controlling these "other" costs

## 3. PROPOSED SOLUTION

Any service of reasonable size—even the fairly small one we recently led [14]—has data centers all over the world. Expanding capacity in, say, Paris can be expensive. The machines have to be shipped to Paris, cleared through customs, and delivered to the data center. Then they must be racked, installed in the data center, burned-in, certified, and labeled for asset tracking. The process requires considerable local skilled labor in addition to the cost of ordering and shipping the individual systems. And servicing failed machines requires someone be in the data center, increasing both costs and administrative errors.

The proposed solution is to no longer build and ship single systems or even racks of systems. Instead, we ship macro-modules consisting of a thousand or more systems. Each module is built in a 20-foot standard shipping container (Figure 1), configured, and burned in, and is delivered as a fully operational module with full power and networking in a ready to run no-service-required package. All that needs to be done upon delivery is provide power, networking, and chilled water.

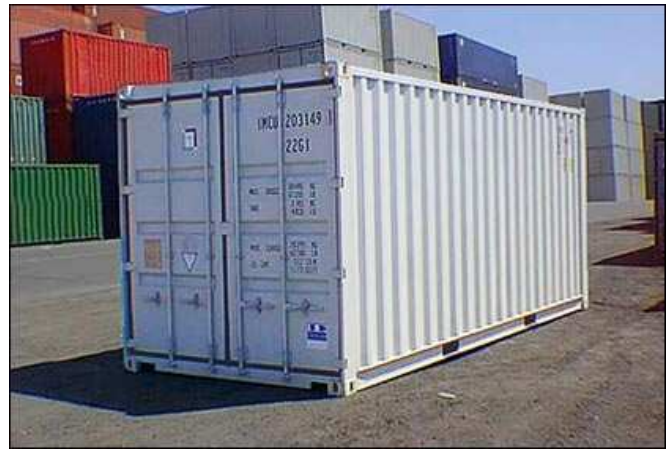

**Figure 1: 20-foot ISO 668 Shipping Container**

The most common concern when this proposal is first discussed is that different applications have very different hardware requirements. Internet-scale services, however, typically choose one or a small number of commodity hardware building-blocks and use these components repeatedly throughout the service. For example, MSN Search uses a two-way x64 system as its core building block [19]. Google was last reported to use a two-way 32-bit Intel-based commodity system as their core compute element [17]. Morgan Stanley reported in March 2006 that Google had moved to 64-bit AMD Opteron[1].

The only way that clusters of 10,000 to 100,000 nodes can be employed cost effectively is to automate all software administration. Those services that have adopted this highly distributed architecture, have automated most aspects of software administration including installation, upgrade, problem detection, and the majority of problem correction. Examples of this approach include Google and Windows Live. After mundane software administrative tasks have been automated and the data center has moved to a full lights-out model, what remains unautomated are the hardware administrative tasks. These include installation, configuration, problem determination, repair, and replacement. This proposal aims at reducing or eliminating these hardware administration tasks.

---

[1] http://www.internetnews.com/ent-news/article.php/3588891.

## 3.1 Shipping Container

Shipping containers are ideal for the proposed solution: they are relatively inexpensive and environmentally robust. They can be purchased new for $1,950 each, while remanufactured units range around $1,500 [11]. The units are designed to successfully transport delicate goods in extreme conditions and routinely spend weeks at a time on the decks of cargo ships in rough ocean conditions and survive severe storms. More importantly, they are recognized world-wide: every intermodal dispatch center has appropriate cargo handling equipment, a heavy duty fork lift is able to safely move them, and they can be placed and fit in almost anywhere. The container can be cost-effectively shipped over the highway or across oceans. Final delivery to a data center is simple using rail or commercial trucking. The container can be placed in any secure location with network capacity, chilled water, and power and it will run without hardware maintenance for its service life.

Many large Internet properties have already taken a step in this direction to reduce some of these overheads. They now order full racks of equipment from the manufacturer, rather than individually packaged systems. Our proposal takes this progression one step further in that the systems are built out fully in the shipping container with full networking support and cooling. The systems don't have to be individually packaged and then placed in a shipping container for transport. And, upon delivery, they don't need to be unpacked and installed in the data center. The container is simply attached to network, chilled water, and power.

Each module includes networking gear, compute nodes, persistent storage, etc. The modules are self-contained with enough redundancy that, as parts fail, surviving nodes continue to support the load. The management model is very similar to the standard reboot, re-image, and replace model used by most Internet-scale services. The only difference is that we don't replace individual failed systems. In this modified model, the constituent components are never serviced and the entire module just slowly degrades over time as more and more systems suffer non-recoverable hardware errors. Even with 50 unrecoverable hardware failures, a 1,000 system module is still operating with 95% of its original design capacity. The principle requirement is that software applications implement enough redundancy so that individual node failures don't negatively impact overall service availability.

This application redundancy problem is well-understood for stateless processing nodes. Most Internet-scale systems operate this way. A somewhat more difficult problem is handling stateful applications, those that save state to disk or depend upon non-recoverable inter-call state. One solution is to write the persistent state to multiple redundant systems that aren't sharing single points of failure. The Google GFS [7] is perhaps the most well-known example of this common, high-scale design pattern.

Unfortunately, allocating redundant application state requires physical network and hardware topology knowledge in addition to a considerable investment in software. The redundant copies, for example, cannot be stored in the same rack, on the same switch, or in the same container. The approach proposed here doesn't change what needs to be done in this dimension. What we are offering here are compute and storage resources in large, uniform slices, delivered and installed in a rugged package. This neither makes the software challenge easier nor more difficult.

At the end of its service life, the container is returned to the supplier for recycling. This service life will typically be 3 years, although we suspect that this may stretch out in some cases to 5 years since little motivation exists to be on the latest technology. During recycling, the container is reloaded with current generation parts. Parts that can't be re-used are recycled. This model brings several advantages: 1) on delivery systems don't need to be unpacked and racked, 2) during operation systems aren't serviced, and 3) at end-of-service-life, the entire unit is shipped back to the manufacturer for rebuild and recycling without requiring unracking & repackaging to ship.

## 3.2 Manufacturing Production Savings

The broad use of the container as a module of data center growth allows substantial economies of scale. For example, robotic assembly/disassembly techniques can be used to reduce manufacturing costs. And, with no provision for field maintenance, the systems can be densely packed without a service aisle. The manufacturer doesn't need to provide training and parts stocking for field maintenance, and error-prone and expensive field servicing errors are avoided.

A further advantage is a container can be treated as a single unit for FCC compliance certification. Similarly, when importing, they are treated as a single unit rather than over 1,000 discrete systems.

## 3.3 Data Center Location Flexibility & Portability

A strong economic argument can be made to consolidate service delivery into a small number of very large data centers. Many factors, however, make this difficult to achieve in practice. First, geo-redundancy is often required. Second, many jurisdictions either require that the data be kept locally or have restrictions on where it can be processed. Some groups and companies believe their data should never be inside the borders of the USA due to concerns about the scope of the USA Patriot Act and related legislation. Similar privacy, liability, or intellectual property concerns apply in other nations. In many cases, these concerns may not be practical as much of the data already flows on the open Internet and quite likely already flows through many foreign jurisdictions. And, even in the cases where it is practical to restrict where the data is stored, broad agreement has not yet emerged on the efficacy of these restrictions.

Nonetheless, many companies are making business decisions on this basis and place restrictions on where their data can be stored. This motivates service providers to use regional data centers, as having them may help to attract more business. Many service providers choose to employ a broad world-wide network of data centers, even though they may not be as efficient to run. The low cost container-based model helps make smaller data centers in remote locations more affordable.

A shipping container is a weatherproof housing for both computation and storage. A "data center," therefore, no longer needs to have the large rack rooms with raised floors that have been their defining feature for years. A central building is still needed to house security, power, networking, and cooling equipment. But the containers can safely be stored outside. The only requirement is a secured, fenced, paved area to place the containers around the central facilities building. The containers can be stacked 3 to 5 high (with support they are packed 7 high on ships) allowing high-density data centers at low build-out costs.

Data centers built using these macro-modules are not only cheap to construct but they are also cheap to move. In the event that network bandwidth is available at lower cost in a different location, or superior tax advantages can be found in a different jurisdiction, or the capacity is no longer needed in this location, the entire data center can be trucked elsewhere. The fixed assets are just a central services building and a fenced compound rather than a $150M facility that must be sold or dismantled.

Data Center design and construction can be slow: a typical 15-megawatt facility takes over 24 months to build. There may be room to accelerate this process using this modular data center approach. Certainly when expanding an existing data center, this approach avoids building permit requirements and, currently, doesn't incur the tax cost of building floor space.

### 3.4 Administrative Savings
On-site hardware service can be expensive in that skilled service personnel must be available at each data center. Our proposal avoids many of these costs. Staffing a data center with full time service technicians is seldom cost-effective unless the facility is very large. Most services contract this work out and can spend 25% of the system's price over a 3-year service life [14]. Even more important than the cost savings, however, are the errors avoided by not having service personal in the data center. Aaron Brown reports that human administrative error causes 20% to 50% of system outages [3]. An increase in overall service availability is also an important gain and perhaps eclipses the cost savings.

The only on-site service required by our proposed approach is installation and management of central power, cooling, networking, and security. Installation only requires network, cooling, and power, so few local skills are needed.

### 3.5 System & Power Density
Typical modern data centers support a power density of approximately 100 watts/sq foot. Some run as high as 350 to 600 watts/sq foot. This relatively low power density often prevents a data center from being fully populated. Even where sufficient power is available, heat density often becomes a problem. Urs Hoelzle explains that Google uses a 1U system design in their data centers and are not considering higher density configurations due to the inability of most data centers to supply the needed power and cooling [8].

For example, Oak Ridge National Lab [13] modeled the cooling requirements for 35 kW racks. The airflow needed to support this configuration is 222,000 CFM, assuming a 6-foot square cooling duct. Their study concluded that the required number of Computer Room Air Conditioners (CRACs) consumes as much space as the systems themselves, not considering the space required for airflow and walkways around the racks. In other words, much less than 50% of the data center floor space can be productively occupied.

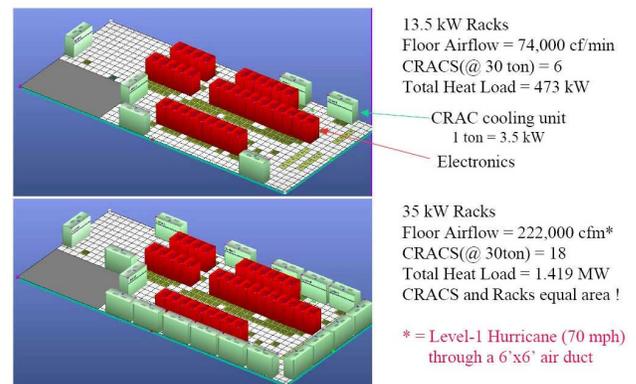

**Figure 2: Air-Cooled Data Centers (Oak Ridge Natl. Lab)**

The Oak Ridge National Lab example shows the inefficiencies of air-to-water cooling. In these configurations, the cooling plant chills coolant that is circulated through CRACs in the machine room. The CRAC cools the air that is circulated through the racks to take off waste heat. This is the approach used in the vast majority of data centers today.

We know from mainframe examples that greater system densities can be supported by direct liquid cooling – fluids have much higher specific heat than air. This is one of those not-so-new discoveries, but the solution seems to have been forgotten. Liquid cooling was mainstream in the 1960s and, for example, IBM used liquid cooling in the mid 80's in the 308x and 309x series Thermal Conduction Modules [9]. A recent storage system built by IBM research shows what heat densities are possible with direct liquid cooling. This system, the IBM Ice Cube, packs 324 disks into a 22.5 inch cube [13]. Liquid-cooled systems racks are now available commercially as well [18]. But liquid cooling hasn't yet been broadly accepted by the rack-and-stack Internet server farms due to complexity and service risk.

The macro-module containers employ direct liquid cooling to eliminate the space requirements for CRAC units. Also, no space is required for human service or for high volume airflow. As a result, the system density can be much higher than is possible with conventional air-cooled racks. The only drawback of direct liquid cooling is the risk of spilling liquid and damaging systems. However, in this service-free approach, the liquid lines are never opened inside the container once sealed at the factory.

## 4. RELATED WORK

As is often the case in data center design, the early innovations around shipping containers first appeared in the telecom industry. Telecom companies have long needed to deliver electronic communications equipment quickly and securely to war zones, construction sites, natural disaster sites, and other locations that lack sufficient interior building space to house the equipment. As a consequence, ISO 668 shipping containers are a common choice, and are part of the price list at companies such as Nortel Networks [16]. The ability to build out a configuration at the manufacturing facility and deliver it quickly and cheaply in a weatherproof steel box is compelling. These units can also be used to expand the telecommunications facilities where floor space is not available, by placing them in parking lots or on roofs.

This low-cost and nimble deployment technique allowed by containerizing is exactly what drove the large power generation companies to also use containers as both the shipping vehicle and the installed housing for some large generator units. Containers delivering 2 megawatts or more are available from several vendors in this form factor [4][6].

Brewster Kahle was an early innovator in applying the containerized transport and delivery approach to data storage as part of the Internet Archive project[2]. Kahle proposed and built the Petabox [12] which is a storage subsystem supporting a petabyte of storage that could be efficiently deployed and shipped anywhere in the world.

Rackable Systems has a macro-module based upon a high boy (9 feet 6 inch rather than the standard 8-foot high) 40 foot shipping container (Figure 3). Their design houses 1,152 systems with a standard-width data center walkway for ease of service while maintaining remarkably high system density.

---

[2] Internet Archive Project http://www.archive.org/index.php.

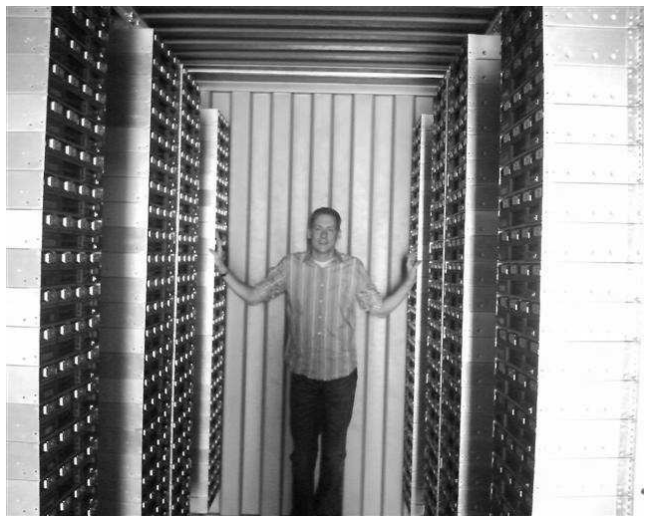

**Figure 3: Rackable Systems Data Center in a Box**

Rackable focuses first and foremost on power and cooling in order to 1) achieve power densities as high as 750 watts/sq ft and 2) achieve cooling power savings approaching 30%. Some of the cooling savings are achieved through improved airflow control in the tight confines of the shipping container and by controlling both fan speed and cooling water flow based upon heat load (Figure 4). Their design scales down well so they are able to supply populated 20 foot units as well.

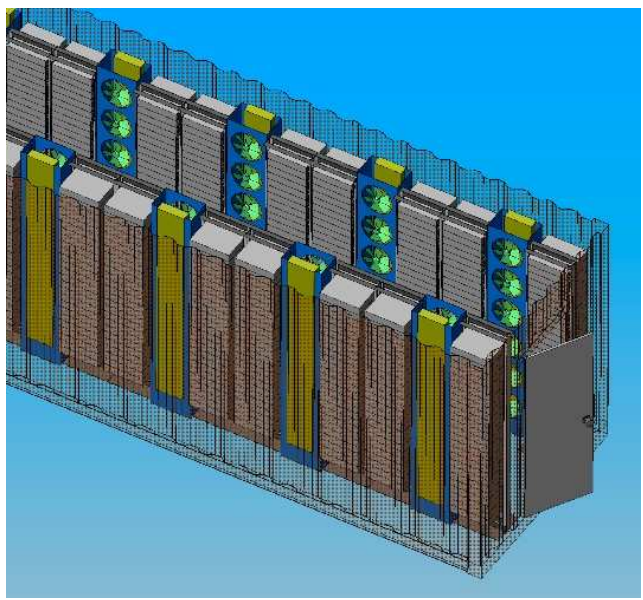

**Figure 4: Rackable Systems Container Cooling Design**

Although the Rackable design hasn't yet been publically announced, they do have a fully populated container is in use at a customer site.

Sun Microsystems also recently announced a macro-module design that they plan to have ready in the summer of 2007. Their approach has many similarities to the one described here (Figure 5). The current Sun prototype is based upon a 20-foot shipping container with 242 systems. This is roughly half the density of the Rackable design, with its 1,152 systems in a 40-foot container.

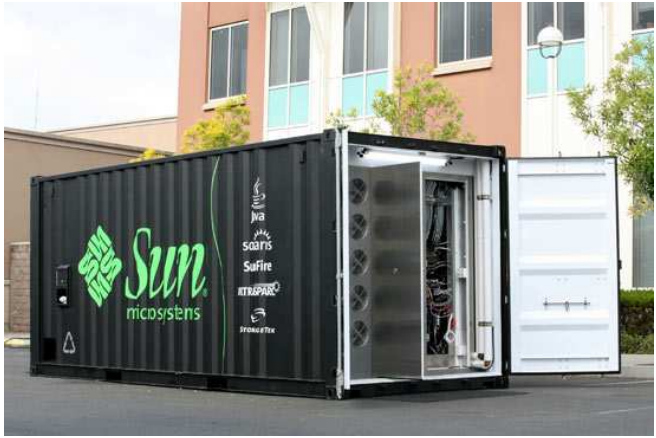

**Figure 5: Sun Microsystems Black Box**

Our approach builds on the work described above in the Telco, power generation, and data storage domains and applies it to modular data center construction and/or extensions. Google may also have related work underway at their California headquarters, but details never been released. This work remains unconfirmed and only reported by an author publishing under the pseudonym Robert X. Cringley [5].

## 5. DESIGN IMPLICATIONS AND FUTURE WORK

The recent work of Rackable Systems and Sun Microsystems suggests that this container-based systems approach may be commercially feasible. Much hardware and software work remains to be done in refining the data center design point. Specifically, are the non-field-serviceable containers proposed in this paper a practical design point? Basically, can a container be assembled at the factory with enough redundancy to run for the full 3-year plus amortization cycle without requiring service? If the non-field serviceable design approach is cost effective, then much higher densities are possible since aisle space is no longer needed and more tightly integrated cooling system designs are practical. Generally, system service constraints make direct liquid cooling difficult to use in a field-serviceable design. Rackable is taking a step in this direction by placing mini-CRAC units beside each rack. However, they still use air-to-water cooling in their approach.

Several vendors currently offer more power-efficient designs. Rather than having an inefficient switching power supply with each system, a rectifier/transformer unit is provided at the rack level and DC power is distributed to all systems within the rack. To avoid losses on the way to the rack, high voltage AC is distributed, with 480VAC being a common choice. Using efficient DC transformers on each system coupled with high efficiency rack-level AC to DC rectifier/transformers yields significant power savings. In addition to being more efficient, this approach has proven to be more reliable in field usage as well.

The design of bringing high voltage AC to the rack and then distributing low voltage DC to each system has drawbacks. Either DC losses to the system must be tolerated or the size of the bus bars carrying the DC must be increased. Using higher voltage DC distribution on the rack has some potential gain but, as the voltage climbs, so does the risk to operations personal. If systems can only be safely moved by electricians, costs escalate to the point of consuming any potential savings. In a non-field-serviceable container, however, higher voltage levels could be employed without this downside cost. What power distribution innovations are possible if the field service cost constraint is removed?

In the design described here, the data center building block is a weatherproof shipping container and these can be stacked 3 to 5 high. What is the ideal data center design with this model? Can we go with a small central networking, power, cooling, and security facility surrounded by stacked containers? Or is it cost-justified to keep the containers in a building? Current data center design theory favors the 10 to 20-megawatt range as the ideal facility size. How does the modular data center change this design point?

Power distribution and equipment costs exceed 40% of the cost of a typical data center, while the building itself costs just over 15%[3]. As a result, a difficult trade-off must be made when selecting power density. If the data center is provisioned to support very high power densities (high power to floor space ratio) the risk is that some of the power will go unused if the actual racks that are installed consume less power/square foot than the data center design point. This would waste expensive power equipment. On the other hand, if the data center is provisioned with a lower power density, the risk is that floor space will be wasted. Wasted floor space is, however, much cheaper than wasted power capacity. Because this trade-off is hard to get exactly correct, most facilities err towards provisioning to lower power densities since a mistake in this direction is much less costly than providing more power than can be used. As a consequence, most facilities are power bound with much wasted floor space. This is an unnecessary cost that also negatively impacts cooling and power distribution efficiency. In a modular design, we can easily adjust data center "floor space" to match the power supply and distribution capabilities of the center. How does this reduction in constraints impact data center design?

The modular data center opens up the possibility of massively distributed service fabrics. Rather than concentrate all the resources in a small number of large data centers with 10 to 20-megawatt power appetites and prodigious networking requirements, can we distribute the computing closer to where it is actually used? Are massively distributed data center designs less network hungry and can sufficient cost savings be found on this design point to justify the increased complexity of managing a more distributed service fabric?

---

[3] Windows Live Operations cost data.

As mentioned in Section 3, the software challenges of building a highly reliable service upon redundant commodity hardware components remain substantial and the work described here neither eases nor worsens that complexity. Fortunately, most Internet-scale services are built using this software design point. However, the challenges of building distributed systems in the 10,000 to 100,000 node range remain significant and, those that are doing it most successfully haven't published extensively. Further research in this area is needed.

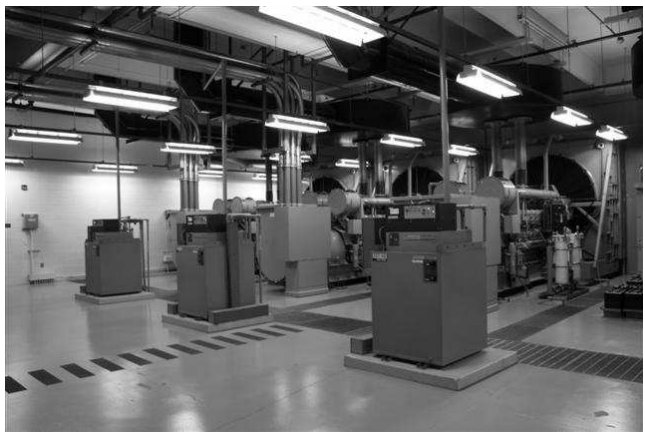

**Figure 6: Microsoft Blue Ridge Generator Room**

Current data center designs are heavily redundant and it is typical to have 10 or more 2.5 megawatt or larger diesel generators at a data center to maintain operation in the event of a power failure (Figure 6). These generators are both expensive to purchase and service intensive. With a large number of modular data centers, would it make sense to not spend on the power redundancy and instead have other data centers pick up the load when a data center fails or is brought down for service? Can we write applications sufficiently tolerant of latency such that we can move load freely between data centers? When does this make sense and what software architectures work best in this environment?

In the electrical power generation field, healthy leasing businesses have been built around large diesel power generators housed in shipping containers as portable power. These are heavily used in war zones, natural disasters command centers, and in allowing permanent facilities to be brought off line for maintenance. The same may be possible for compute and storage capacity with leasing businesses being built around delivering compute and storage quickly efficiently to locations with poor infrastructure or to meet seasonal or unexpected requirements at existing facilities.

## 6. CONCLUSION

In this paper, we propose a new data center macro-module based upon 20x8x8-foot shipping containers. We show savings in agility, capitalization, deployment, management, maintenance, availability, and in recycling at module end of life. In addition, each container is non-serviceable and, as components fail, redundant capacity takes over and the module remains operational. This substantially reduces administrative costs and also eliminates 20% to 50% of the downtime attributed to administrative error in conventional data centers. Because these modules are directly water-cooled, much higher system densities can be achieved, further reducing space requirements. Rather than needing a large data center building to house all the equipment, the systems are shipped and installed in standard, weatherproof shipping containers. We need only a small administrative building for security, power, cooling, networking and power. The entire data center can be easily relocated by truck, rail or ship. This architecture transforms data centers from static and costly behemoths into inexpensive and portable lightweights.

## 7. AKNOWLEDGEMENTS

I thank the anonymous CIDR reviewers and appreciate their input. Xiaodong Huang helped formulate the early ideas during brainstorming sessions on data center efficiency, automation, and the role of hardware innovation in the data center. I thank Matt Gambardella and Giovanni Coglitore of Rackable Systems for contributing unreleased technical material, pictures, and empirical data from a customer-installed beta configuration. Atul Adya, Andrew Cencini, Sam Christi, Filo D'Souza, Jim Gray, Jennifer Hamilton, David Nichols, Pat Selinger, and David Treadwell all contributed valuable review comments.